\documentclass[aps,prd,showpacs,showkeys]{revtex4}

\usepackage{amssymb,amsmath}
\usepackage[pdftex]{graphicx}
\DeclareGraphicsExtensions{.pdf}
\usepackage{subfig}
\usepackage{bm}
\usepackage{color}
\usepackage{paralist}

\begin{document}

\title{Quantitative Analysis of LISA Pathfinder Test Mass Noise}

\author{Luigi Ferraioli}
\email{luigi@science.unitn.it}
\affiliation{University of Trento and INFN, via Sommarive 14, 38123 Povo (Trento), Italy}

\author{Martin Hewitson}
\affiliation{Albert-Einstein-Institut, Max-Planck-Institut fuer Gravitationsphysik und Universit\"{a}t Hannover, Callinstr. 38, 30167 Hannover, Germany}

\author{Giuseppe Congedo}
\affiliation{University of Trento and INFN, via Sommarive 14, 38123 Povo (Trento), Italy}

\author{Miquel Nofrarias}
\affiliation{Institut de Ci\`encies de l'Espai, (CSIC-IEEC), Facultat de Ci\`encies,
Campus UAB, Torre C-5, 08193 Bellaterra, Spain}

\author{Mauro Hueller}
\affiliation{University of Trento and INFN, via Sommarive 14, 38123 Povo (Trento), Italy}

\author{Michele Armano}
\affiliation{SRE-OD ESAC, European Space Agency, Camino bajo del Castillo s/n, Urbanizaci\'{o}n Villafranca del Castillo, Villanueva de la Ca{\~n}ada, 28692 Madrid, Spain}

\author{Stefano Vitale}
\affiliation{University of Trento and INFN, via Sommarive 14, 38123 Povo (Trento), Italy}

\begin{abstract}

LISA Pathfinder (LPF) is a mission aiming to test the critical technology
for the forthcoming space-based gravitational wave detectors. The main scientific objective
of the LPF mission is to demonstrate test-masses free-falling with residual accelerations
below $3 \times 10^{-14} \textnormal{ m} \textnormal{ s}^{-2} / \sqrt{\textnormal{Hz}}$
at $1$ mHz. Reaching such an ambitious target will require a significant
amount of system optimisation and characterisation, which will in turn require
accurate and quantitative noise analysis procedures. In this paper we discuss two main problems associated with the analysis of the data from LPF:
\begin{inparaenum}[\itshape  i\upshape)]
\item Excess noise detection and
\item Noise parameter identification.
\end{inparaenum}
The mission is focused on the low frequency region ($\left[0.1, 10\right]$ mHz) of the available signal spectrum. In such a region the signal is dominated by the force noise acting on test masses. At the same time, the mission duration is limited to $90$ days and typical data segments will be $24$ hours in length.
Considering those constraints, noise analysis is expected to deal with a limited amount of non-Gaussian data, since the spectrum statistics will be far from Gaussian and the lowest available frequency is limited by the data length. 
In this paper we analyze the details of the expected statistics for spectral data and develop two suitable excess noise estimators. One is based on the statistical properties of the integrated spectrum, the other is based on Kolmogorov-Smirnov test. The sensitivity of the estimators is discussed theoretically for independent data, then the algorithms are tested on LPF synthetic data. The test on realistic LPF data allows the effect of spectral data correlations on the efficiency of the different noise excess estimators to be highlighted. It also reveals the versatility of the Kolmogorov-Smirnov approach, which can be adapted to provide reasonable results on correlated data from a modified version of the standard equations for the inversion of the test statistic.
Closely related to excess noise detection, the problem of noise parameter identification in non-Gaussian data is approached in two ways. One procedure is based on maximum likelihood estimator and another is based on the Kolmogorov-Smirnov goodness of fit estimator. Both approaches provide unbiased and accurate results for noise parameter estimation and demonstrate superior performance with respect to standard weighted least-squares and Huber's norm. We also discuss the advantages of the Kolmogorov-Smirnov formalism for the estimation of confidence intervals of parameter values in correlated data.

\end{abstract}



\maketitle

\section{Introduction}
\label{intro}

LISA Pathfinder (LPF), a European Space Agency mission, will be used to characterize
and analyze all possible sources of disturbance which perturb free-falling test
masses from their geodesic motion. The system is composed of a single spacecraft
(SC) enclosing a scientific payload, the LISA Technology Package (LTP), which
is composed of two test masses (TMs) whose position is sensed by an interferometer.
The spacecraft cannot simultaneously follow both masses, therefore the trajectory
of only one test mass is used as a drag-free reference along the measurement
axis. In order to prevent the trajectories of the test-masses from diverging, the
second test mass is capacitively actuated to follow the first (free-falling) TM.
In the main science operating mode, the first interferometer
channel measures the displacement of the SC relative to the free-falling TM.
The second interferometer channel (the differential channel) measures the relative
displacement between the two TMs.

LPF is a controlled system which can only be fully assessed during flight operation,
therefore a considerable number of experiments will be devoted to the identification
of the details of the dynamics of the system. A dynamical model of  LPF
is built in advance on the basis of physical considerations and from the results of test campaigns. The dynamical model is parametric so that it can be updated
on the basis of the experiments that will be conducted during mission operations.
 The overall aim of the process is to reach the best free-fall quality (below
$3 \times 10^{-14} \textnormal{ m} \textnormal{ s}^{-2} / \sqrt{\textnormal{Hz}}$
at frequencies around $1$ mHz) in a step-by-step procedure in which the
result of the previous experiment is used to adjust the detailed configuration of
the following experiments \cite{LTPVitale2011,LTPMartin2011,LTPPaul2011,ArmanoCQG2009}.

Such a demanding program requires daily analysis of the instrument signals constrained by two major factors:
\begin{inparaenum}[\itshape  i\upshape)]
	\item the amount of available data is tightly constrained by LTP mission duration ($90$ days), the telemetry bandwidth, and the length of each data segment (typically $24$ h);
	\item the scientific interest is mainly focused on the analysis of noise sources which act directly on the TMs since that should provide a baseline reference for the forthcoming space-based gravitational waves observatories \cite{LISAYellowBook, Cornish2005, DECIGO, ASTROD}.
\end{inparaenum}
The direct forces on the TMs are expected to dominate the  instrument output in the frequency range $\left[0.1, 10\right]$ mHz. Sample power spectra are typically calculated with Welch's averaging periodogram method (WOSA) \cite{Percival}. In order to keep enough frequency resolution at low frequencies, the sample power spectra can not be averaged more than few times (we average $4$ times in the present paper), this results in highly non-Gaussian data for which we are developing dedicated techniques.
In particular the paper aims to propose a solution for two major data analysis challenges
encountered in LPF:
\begin{inparaenum}[\itshape  i\upshape)]
\item Different measurements of the same physical quantity can exhibit different
noise content if they are performed under slightly different environmental conditions.
The objective of LISA Pathfinder data analysis during operations will
be to discover such differences, understand their origin and adjust spacecraft physical
parameters accordingly. Such a problem requires reliable excess noise
detection procedures which have to be based on solid statistical considerations;
\item Along with the demonstration of unprecedented test-mass free-fall, LPF will
provide a model for the expected test mass force noise for future space-based gravitational wave detectors. In order to do this, we need to be able to match an analytical model to a noisy power spectral density measurement. The quality of the match must be statistically
quantified.
\end{inparaenum}
Both data analysis problems deal with sample spectra and
the corresponding statistical properties. 
Section \ref{sect.noisexcess} reports on the properties of different experimental procedures for the detection of noise excess. In particular we considered two cases where the noise excess is evaluated with respect to reference data or a reference model. The accuracy of the methods is theoretically analyzed for the case of broadband and band-limited excess noise.
In Section \ref{sect.noisemodident} the problem of noise parameter estimation for non-Gaussian data is explored and an algorithm based on maximum likelihood is derived. In parallel, the application of the Kolmogorov-Smirnov formalism to the construction of a goodness of fit estimator is discussed. 
Section \ref{sect.nonstatnoise} reports briefly on the extension
of the analysis procedures to the case of non-stationary noise and time-frequency
investigations. 
In Section \ref{sect.testofperformance} we provide an application of the developed algorithms to synthetic data with LPF-like qualities. This allows us to shed light on the effects of data correlations on the accuracy of the developed excess noise estimators.
The analysis procedures are available as MATLAB tools in the framework
of the LTPDA Toolbox \cite{HewitsonCQG2009, matlab, toolbox}; an object oriented
MATLAB Toolbox for advanced data analysis.



\section{Quantitative detection of noise power variations}\label{sect.noisexcess}

The problem of the detection of noise power variation in consecutive measurements can be formulated in two different ways.
\begin{inparaenum}[\itshape  i\upshape)]
\item Two different measurements of the power are compared;
\item The different measurements of the power are compared with a reference model.
\end{inparaenum}
The problem in the first case is of general character and can be applied to a wide range of experiments, the second case, on the other hand, assumes that a reference model for the noise power is known and data must be compared against the given model in order to understand if the system is performing under known conditions. The latter case is likely to be the scenario for LPF operations.
The quantity typically used for the detection of noise power variations is the total energy content of the data series. It is defined as $\mathcal{E} = T \sum_i{\left|x_i\right|^2}$. Unfortunately, $\mathcal{E}$, provides a poor estimator for two reasons:
\begin{inparaenum}[\itshape  i\upshape)]
\item As soon as the data series $x_0,\ldots,x_{N-1}$ departs from zero mean Gaussian white noise, the statistic of $\mathcal{E}$ becomes ill-defined and the definition of a confidence interval becomes cumbersome;
\item $\mathcal{E}$ provides global information as it is not sensitive to noise changes in a given frequency band.
\end{inparaenum}
While the first problem could be overcome (without little difficulty) by a numerical identification of the expected statistic, the second problem suggests that a spectral based estimator would provide supplementary information which could be fundamental in discriminating different noise sources.

\subsection{Detection of noise variations with a model}\label{subsect:noisexcsmodel}

In the case that a reference model is available, the detection of excess noise in the spectral domain can be effectively implemented with a test on the normalized Welch's overlapped segment averaging (WOSA) spectrum $R_{\text{WOSA}}\left(f_k\right)$ defined in equation (\ref{eqn.normwosa}).
In particular, the test can pursue two different philosophies of which one aims to test a global scalar indicator of the properties of the data and another aims to test the details of the statistical distribution of the data.

A sensible estimator for the first approach is provided by the integral of the normalized spectrum which, in the discrete case, can be written:
\begin{equation}\label{eqn:IRdef}
	\mathrm{IR} = \sum_{k=1}^{N_f}{R_{\text{WOSA}}\left(f_k\right)}.
\end{equation}

In the simplifying assumption of independent spectral data, the statistic of each element of $R_{\text{WOSA}}\left(f_k\right)$ is described by a gamma distribution as defined by equation (\ref{eqn.probdistwosa}) with $\delta = 1/N_s$ and $h = N_s$. The sum over the different values at the frequencies $f_k$ is still a gamma distribution with $\delta = 1/N_s$ and $h = N_s N_f$. The expectation value for $\mathrm{IR}$ is easily obtained as $E\left[\mathrm{IR}\right] = \delta h = N_f$.

The natural estimator for the second approach is provided by the empirical cumulative distribution function (ECDF). The ECDF for a data series can be defined as $F_N\left(x\right) = z/N$, where $z$ is the number of observations reporting a value less than or equal to $x$. $x$ denotes the values taken by the data, in our case $R_{\text{WOSA}}\left(f_k\right)$ with $k = 1,\ldots,N_f$. The ECDF can be tested against the theoretical expectation provided by equation (\ref{eqn.probdistwosa}). If the model well represents the given sample spectrum then $R_{\text{WOSA}}\left(f_k\right)$ is distributed according to the expected gamma distribution. Alternatively, if the sample spectrum contains a excess noise with respect to the reference model, then the distribution of the normalized WOSA spectrum will be different from the one reported in equation (\ref{eqn.probdistwosa}). The hypothesis that the two distributions are
equal can be tested if a 'distance' between
the ECDF and the theoretical reference, $F_{\Gamma}$, is defined as:
\begin{equation}\label{eqn.ksdistance}
	d_K\left(x\right) = \left|F_N\left(x\right) - F_{\Gamma}\left(x\right)\right|,
\end{equation}

with $d_K\left(x\right)$ assuming values on the interval $\left[0,1\right]$ and $K = N_f$. Kolmogorov found that the statistical properties of 
\begin{equation}\label{eqn.dKdef}
	d_K = \text{max}\left[d_K\left(x\right)\right]
\end{equation}

are independent from the specific distributions under test \cite{Kolmogorov1933, Feller1948}. This property qualifies $d_K$ as an excellent candidate for the construction of a general test for cumulative distribution functions, the limiting statistic for $d_K$ was identified by Kolmogorov himself and then inverted by Smirnov \cite{Smirnov1939,Miller1956} who provided an analytical expression for the calculation of $d_K$ as a function of the significance level.
General details about the Kolmogorov-Smirnov test (KS-test)
are reported in appendix \ref{appendix.KStest}.

It is interesting to calculate the expected sensitivity for the two estimators. $\mathrm{IR}$ is expected to be distributed as a gamma distribution (at least when the model and the data are in agreement), the corresponding cumulative distribution function (CDF) is provided by the incomplete gamma function $\mathcal{P}\left(h,x/\delta\right)$ \cite{AbramoStegun}. The CDF assumes values in the interval $\left[0,1\right]$ as it defines the probability associated with the observation $x$. The inverse of the CDF provides the critical values, $x_{\rho}$, associated with a given probability, $\rho \in \left[0,1\right]$. The confidence range for a probability, $\rho$, for a gamma distributed variable can then be defined by the boundaries $x_{lw} = \mathcal{P}^{-1}\left(\alpha/2\right)$ and $x_{up} =  \mathcal{P}^{-1}\left(1-\alpha/2\right)$, with $\alpha = 1 - \rho$. We assert that the measured sample spectrum is compatible with the reference model if $x_{lw} \leq \mathrm{IR} \leq x_{up}$ for the given probability, $\rho$, or significance level, $\alpha$.

If the noise excess is provided by a scale factor, $\gamma$, which affects the noise on the complete band of frequencies, the expected value for $\mathrm{IR}$ changes to $E\left(\mathrm{IR}\right) = \gamma N_f$. Therefore the detection threshold for $\gamma$ is fixed by the interval $x_{lw}/N_f \leq \gamma \leq x_{up}/N_f$. In other words, the $\mathrm{IR}$ can detect a noise difference with respect to the reference model only if $\gamma < x_{lw}/N_f$ or if $\gamma > x_{up}/N_f$.
If $\gamma$ is non-zero only in a restricted band of frequencies $\left[f_a,f_b\right]$, then the expectation value for $\mathrm{IR}$ changes to $E\left(\mathrm{IR}\right) = N_f - N_{ab} + N_{ab} \gamma$. In this case $\mathrm{IR}$ can detect the noise difference only if $\gamma < \left(x_{lw} - N_f + N_{ab}\right)/N_{ab}$ or $\gamma 
> \left(x_{up} - N_f + N_{ab}\right)/N_{ab}$. $N_{ab}$ is the number of frequency points in the interval $\left[f_a,f_b\right]$.

In the case of the Kolmogorov-Smirnov (KS) estimator, the ECDF of the WOSA normalized spectrum is compared with the theoretical expectation $\mathcal{P}\left(N_s,x N_s\right)$. In the case that the noise excess is simply a constant scale factor, $\gamma$, over the whole frequency band, the expected CDF for the normalized WOSA spectrum is $\mathcal{P}\left(N_s,x N_s/\gamma\right)$. The expected value for the KS test variable is then written as:
\begin{equation}\label{eqn.ksvardetectth}
	d_K\left(\gamma\right) = \mathrm{max}\left|\mathcal{P}\left(N_s,\frac{x N_s}{\gamma}\right)-\mathcal{P}\left(N_s,x N_s\right)\right|.
\end{equation}

Once a significance value $\alpha$ is provided, the corresponding critical value, $d_K\left(\alpha\right)$, can be calculated from the equations for the inversion of the limiting CDF for $d_K$ \cite{Smirnov1939,Miller1956}. The two distributions in equation (\ref{eqn.ksvardetectth}) are incompatible at the given significance level if $d_K\left(\gamma\right) > d_K\left(\alpha\right)$. This defines the detection threshold for $\gamma$. If $\gamma$ is non-zero only in a restricted band of frequencies $\left[f_a,f_b\right]$, then the expected distribution for the normalized WOSA estimator is difficult to calculate. Nevertheless the detection threshold for the KS estimator can be calculated numerically from synthetic data.

\begin{figure}
\includegraphics[width=0.45\textwidth]{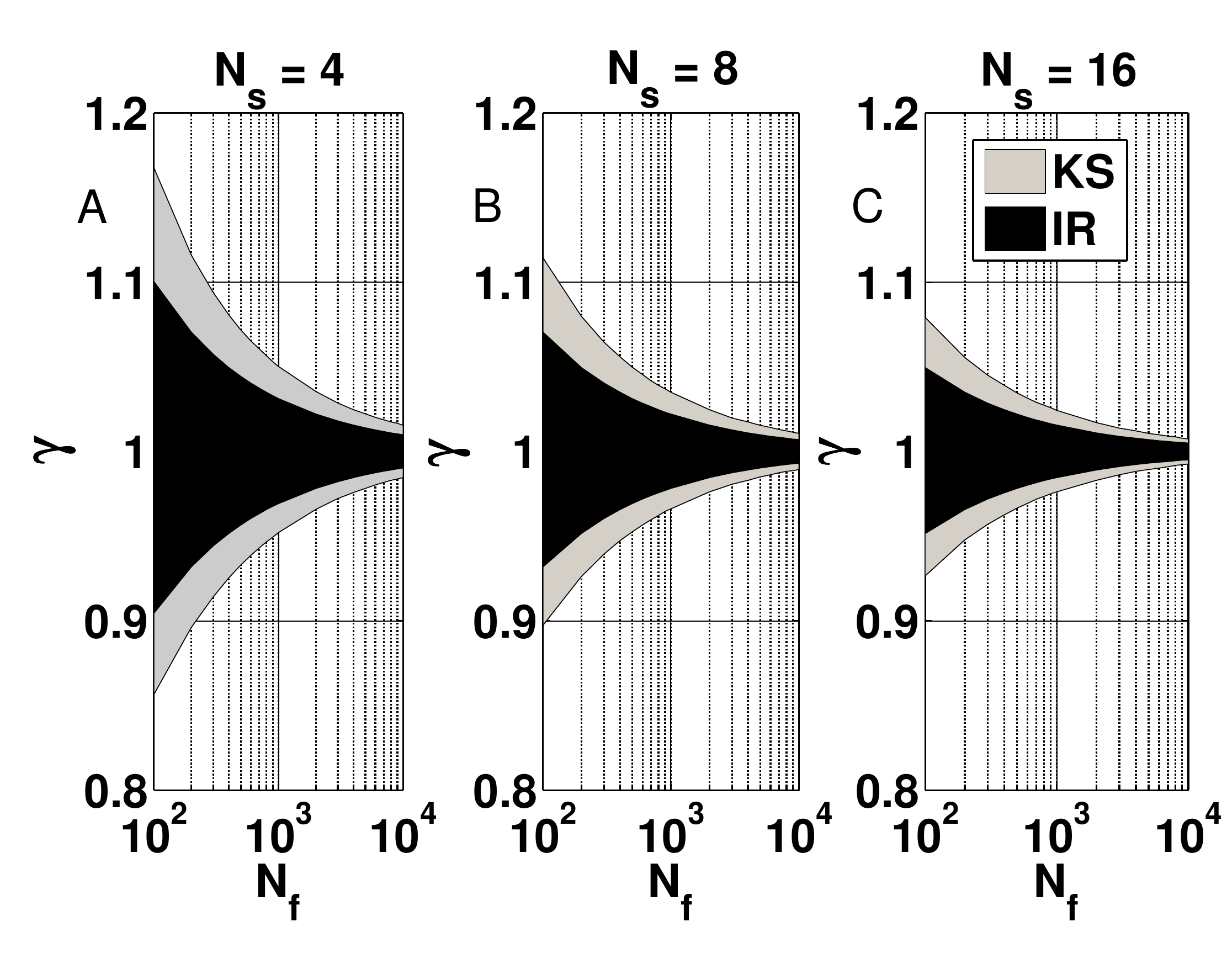}
\caption{Non-detection ranges for $\mathrm{IR}$ and KS estimators. Noise excess is assumed as a constant multiplicative factor $\gamma$ extending along the whole band of frequencies. Darkened intervals define the thresholds for $\gamma$ detection, if $\gamma$ is larger or smaller than the shaded values it can be detected with a confidence of $95 \%$. $N_s$ refers to WOSA averages. $N_f$ is the number of frequency points in the spectrum.}
 \label{fig:fig1}
\end{figure}

\begin{figure}
\includegraphics[width=0.45\textwidth]{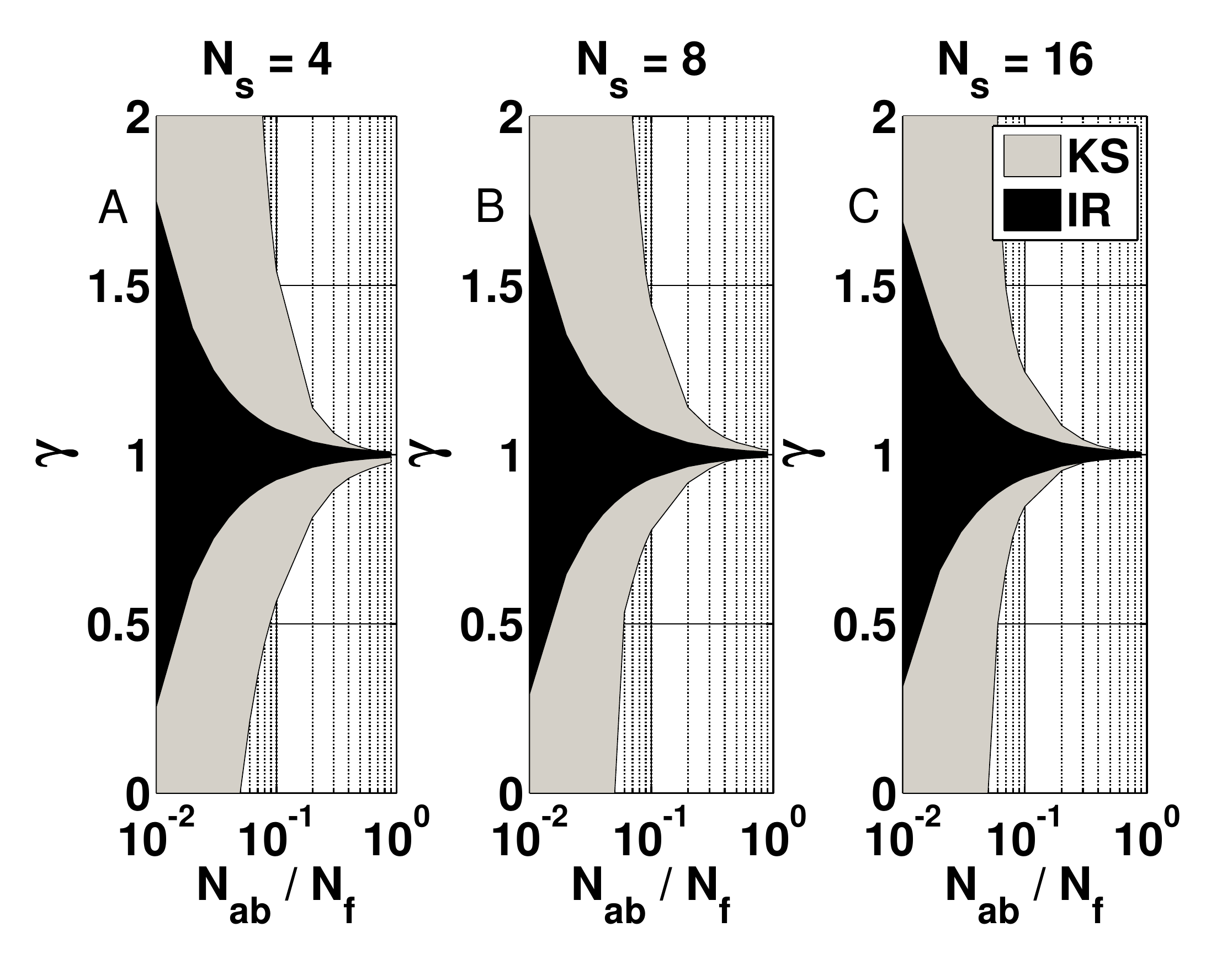}
\caption{Non-detection ranges for $\mathrm{IR}$ and KS estimators in the case that noise excess coefficient $\gamma$ is different from zero in a restricted band $\left[f_a,f_b\right]$. Data are presented as a function of the ratio $N_{ab} / N_f$ where $N_{ab}$ is the number of frequency points in the interval $\left[f_a,f_b\right]$ and $N_f$ is the total number of frequency points in the spectrum. $N_s$ refers to WOSA averages. Confidence level for excess detection is fixed to $95 \%$.}
 \label{fig:fig2}
\end{figure}

In figure \ref{fig:fig1}, the ranges of non-detectability of the KS and $\mathrm{IR}$ estimators are reported as a function of $N_f$ for three different values of the WOSA averages $N_s$. Data refer to the case that the scale factor, $\gamma$, extends over the complete frequency band. As can be clearly seen, the $\mathrm{IR}$ estimator always has a better sensitivity than the KS estimator.
The sensitivity for the case of a band limited excess noise is reported in figure \ref{fig:fig2} as a function of the ratio $N_{ab}/N_f$. As in the previous case, the $\mathrm{IR}$ estimator provides a better sensitivity with respect to the KS estimator. The difference is particularly relevant for $N_{ab}/N_f < 0.07$, below such values the sensitivity of the KS estimator becomes poor. Both in figure \ref{fig:fig1} and in figure \ref{fig:fig2} the confidence level for $\gamma$ detection is fixed at $95 \%$. 

It is worth noting that KS algorithm can be used directly on time series to quantitatively gauging departures from a given distribution (e.g. Gaussian). Once the ECDF for the data is calculated, it can be compared with the expected distribution by calculating $d_K(x)$ from equation \ref{eqn.ksdistance}. Since the distribution of $d_K(x)$ coefficients is known it is straightforward to set a confidence threshold. The procedure can be applied even in presence of correlations thanks to the generalizations discussed in section \ref{subsect.test.excessnoise} and in appendix \ref{appendix.KStest}.

\subsection{Detection of noise variations without a model}

In the case that an excess of noise has to be detected by comparing different measurements the Parseval's theorem \cite{Percival} ($\mathcal{E} = \sum_{k=0}^{N/2}{P\left(f_k\right)}$) suggests that the sum of the elements of a sample spectrum in a given frequency band $\mathcal{E}_{ab} = \sum_{k=a}^{b}{P\left(f_k\right)}$ could provide a sensitive estimator for noise power variations. Such an estimator would be loosely equivalent to the $\mathrm{IR}$ discussed above, except that its statistic is hard to determine in a typical experimental situation. The statistic of $P\left(f_k\right)$ at each frequency, $f_k$, is defined by equation (\ref{eqn.sampspprobdist}). Therefore, in the case of non-white noise, its parameters depend on $S\left(f_k\right)$ and the statistic is different at different frequencies. Thus the statistic of $\mathcal{E}_{ab}$ is not easily known.

An interesting alternative to $\mathcal{E}_{ab}$ is provided by the KS estimator. 
Given two data series $\left\{x_{n_1}\right\}$ and $\left\{x_{n_2}\right\}$ of length $N_1$ and $N_2$ respectively. The hypothesis that their ECDFs have the same limiting cumulative distribution function $F\left(x\right)$ can be tested if a distance in the ECDFs space is defined as:
\begin{equation}\label{eqn.ksdistanceemp}
	d_K\left(x\right) = \left|F_{n_1}\left(x\right) - F_{n_2}\left(x\right)\right|,
\end{equation}

with $d_K\left(x\right)$ defined on the interval $\left[0,1\right]$ and $K = \left(N_1 N_2\right) /\left(N_1 + N_2\right)$ \cite{Feller1948}. Also in this case, the statistical properties of 
\begin{equation}\label{eqn.dKdefemp}
	d_K = \text{max}\left[d_K\left(x\right)\right]
\end{equation}
are independent from the distributions of $x_{n_1}$ and $x_{n_2}$. 
The same equations used in the case of the comparison with a given model can now be used for the inversion of the cumulative statistic of $d_K$
\footnote{As formulated by equation
(\ref{eqn.dKdefemp}), the test is searching for differences in the noise content
of the data series. It does not provide information about which series has the
largest noise content. Such information can be recovered graphically with a distribution
plot (for example, a quantile-quantile plot). Alternatively, a pure excess noise
test for a data series, $x_m$, with respect to a reference series, $x_n$, can be formulated
by substituting $d_K$ with $d^{+}_K = max\left\{F_{n_2}\left(x\right) -
F_{n_1}\left(x\right)\right\}$. The formalism used for the calculation of $d_K$ and
$d^{+}_K$ is the same, therefore the procedures described in this paper can
be applied in either case without any significant modification.}.
Considering the simplifying assumption of independent spectral data, the sensitivity of the KS estimator at a given significance level can be calculated in analogy to what was discussed in the previous paragraph.

\begin{figure}
\includegraphics[width=0.45\textwidth]{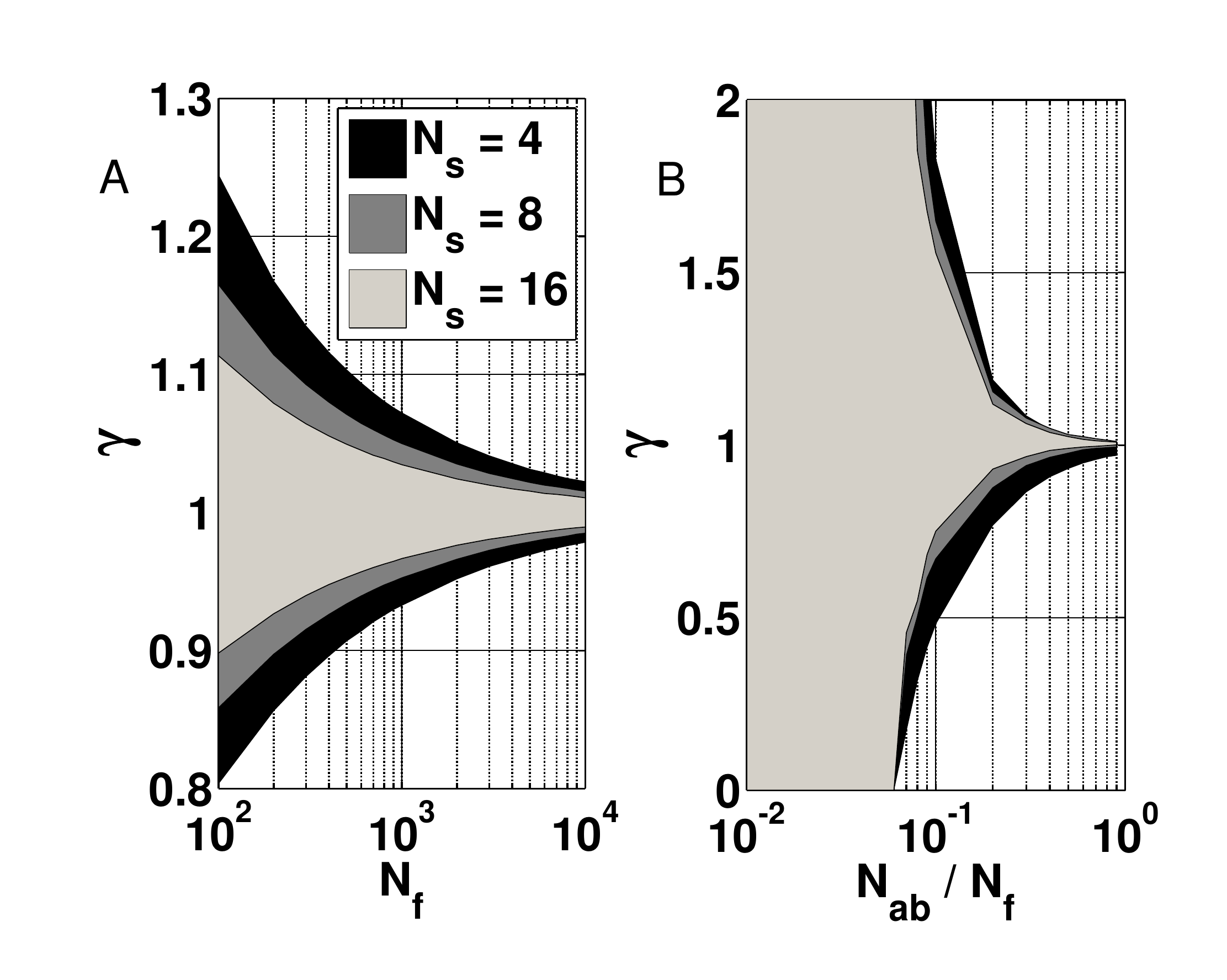}
\caption{Non detection ranges for the KS estimator for the comparison of two sample spectra. A) Noise excess coefficient  $\gamma \neq 0$ on the full frequency band. $N_s$ is the number of WOSA averages. $N_f$ is the number of frequency points in the sample spectrum. B) $\gamma \neq 0$ only in a restricted interval of frequencies $\left[f_a,f_b\right]$. $N_{ab}$ is the number of points in the interval. }
 \label{fig:fig3}
\end{figure}

In figure \ref{fig:fig3} the calculated interval for non-detection is reported for the case that the excess, $\gamma$, is extending over the whole frequency band and for the case that $\gamma \neq 0$ in $\left[f_a,f_b\right]$.


\section{Noise model identification}\label{sect.noisemodident}

Closely related to the problem of excess noise detection, the problem of noise model identification is one of the principal scientific objectives of LPF mission.
Of particular interest is the identification of a model for the force noise acting on the TMs, which can be used as a baseline for the forthcoming space-based gravitational wave observatories.

The main constrains on the identification of force noise on the TMs in LPF are fixed by 
\begin{inparaenum}[\itshape  i\upshape)]
\item the limited data series length, typically $24$ h;
\item the frequency range in which force noise on the TMs is dominating the signal is below $10$ mHz;
\item force noise data are not directly accessible since the system measures and reports TM displacement, requiring that force noise on TMs be reconstructed by a numerical procedure \cite{LISA7Anneke, LISA7Luigi}.
\end{inparaenum}

The result of the combination of the first two constraints is that the number of segment averages in the WOSA procedure for sample spectrum estimation should be taken as low as possible so as to have a reasonable number of frequency points in the range $f \in \left[0.1,10\right]$ mHz.
As a consequence, the distribution of the WOSA spectrum strongly departs from a Gaussian distribution
\footnote{The distribution of the WOSA estimator for the sample spectrum is a gamma distribution, as reported in equation (\ref{eqn.probdistwosa}). Such a distribution tends to a Gaussian distribution as the number of averages increases. The difference between the `true' distribution and the corresponding Gaussian can be quantified by the maximum distance among cumulative distribution functions, $d$. As an example $d = 0.067$ for $N_s = 4$, $d = 0.03$ for $N_s = 16$, $d = 0.019$ for $N_s = 50$, $d = 0.013$ for $N_s = 100$ and $d = 0.0094$ for $N_s = 200$. It is worth noting that since $d$ is a difference in the space of the values of the CDF, it can assume values in $\left[0,1\right]$. As can be seen, the distribution of the WOSA spectrum reasonably approaches a Gaussian only for $N_s$ as large as $200$.}, 
meaning that the classical least-squares minimization procedure for parameter estimation is not well conditioned and a full maximum likelihood procedure is required.

\subsection{Maximum Likelihood Parameter Estimation}\label{subsect.mlikelihhod}

If we replace $S\left(f_k\right)$ in equation (\ref{eqn.normwosa}) with a parametric model for the spectrum, $S\left(f_k;\theta_1,\ldots,\theta_H\right)$, the normalized WOSA spectrum becomes parametric, $R_{\text{WOSA}}\left(f_k;\theta_1,\ldots,\theta_H\right)$, and can be used for the estimation of noise model parameters $\left\{\theta_1,\ldots,\theta_H\right\}$. In this section we develop the likelihood formalism for the simple case that the noise model is a function of a single parameter, $S\left(f_k;\theta\right) = \theta S_0\left(f_k\right)$. This allows us, not only to find a sensible goodness of fit estimator for $R_{\text{WOSA}}\left(f_k;\theta\right)$ that can be used also in the case of multiple parameters, but also to place the excess noise estimator, $\mathrm{IR}$, in a more solid theoretical framework.

Indicating with $\theta_{\text{TRUE}}$ the `true' value for the $\theta$ parameter, $R_{\text{WOSA}}\left(f_k;\theta\right)$ can be rewritten as:
\begin{equation}
\begin{split}
	R_{\text{WOSA}}\left(f_k;\phi\right) & = \phi \frac{P_{\text{WOSA}}\left(f_k\right)}{\theta_{\mathrm{TRUE}} S_0\left(f_k\right)} \\
	& = \phi R_{\text{WOSA}}^{\mathrm{true}}\left(f_k\right).
\end{split}
\end{equation}

Here $\phi = \theta_{\mathrm{true}} / \theta$. Assuming that $\theta_{\mathrm{TRUE}} S_0\left(f_k\right)$ correctly reproduces the expected value for the spectrum, the distribution of $R_{\text{WOSA}}^{\mathrm{true}}\left(f_k\right)$ is reported in equation (\ref{eqn.probdistwosa}) with $\delta = 1/N_s$ and $h = N_s$. The distribution of the samples $R_{\text{WOSA}}\left(f_k;\phi\right)$ is then:
\begin{equation}
	f\left(y;h,\delta,\phi\right) = \frac{e^{-\frac{y}{\phi \delta}} y^{h-1} }{\left(\phi \delta\right)^h \Gamma\left(h\right) }.
\end{equation}

Under the simplifying assumption that the values of $R_{\text{WOSA}}\left(f_k;\phi\right)$ are independent for different $f_k$, the likelihood function can be written as:
\begin{equation}\label{eqn.NormWOSALike}
	\mathcal{L}\left(h,\delta,\phi\right) = \prod_{k} f\left(y_k;h,\delta,\phi\right) \Delta y.
\end{equation}
Here $\Delta y$ is a constant term required to have a finite probability from the probability distribution function $f\left(y_k;h,\delta,\phi\right)$. $y_k$ are observed samples corresponding to $R_{\text{WOSA}}\left(f_k;\phi\right)$. It is typically more convenient to work with the natural logarithm of the likelihood function $l\left(h,\delta,\phi\right) = \ln \mathcal{L}\left(h,\delta,\phi\right)$.
\begin{equation}
	l\left(h,\delta,\phi\right) \sim \left(h-1\right) \sum_k{\ln y_k} - \frac{1}{\phi \delta} \sum_k{y_k} - N_f h \ln \phi.
\end{equation}
$N_f$ is the total number of frequency samples. 
Taking the first derivative with respect to $\phi$ and equating to $0$ we find the maximum likelihood estimator for the parameter $\phi$:
\begin{equation}
	\Lambda = \frac{1}{\phi} \sum_k{R_{\text{WOSA}}\left(f_k;\phi\right)} - N_f.
\end{equation}

The value of $\phi$ at which $\Lambda = 0$ corresponds to the maximum likelihood estimation for the parameter. The $\Lambda$ estimator is unbiased, since, remembering that $\phi = \theta_{\mathrm{true}} / \theta$, it can be verified that $\lim_{\theta \rightarrow \theta_{\mathrm{true}}} E\left[\Lambda\right] = 0$. For the practical purpose we can find the zero crossing of the reduced estimator
\begin{equation}
	\tilde{\Lambda} = \sum_k{R_{\text{WOSA}}\left(f_k;\theta\right)} - N_f,
\end{equation}
which is crossing zero at the same $\theta$ value of $\Lambda$, since $\phi \rightarrow 1$ when $\theta \rightarrow \theta_{\mathrm{true}}$.
It is worth noting that $\tilde{\Lambda}$ is practically the $\mathrm{IR}$ excess noise estimator with the expectation value $N_f$ subtracted.
It is then worth noting that $\tilde{\Lambda}$ can be used not only in the simple case of one parameter but it can also be applied in the general case of a model $S\left(f_k;\theta_1,\ldots,\theta_H\right)$ since the condition $\tilde{\Lambda} \rightarrow 0$ when $\left\{\theta_1,\ldots,\theta_H\right\} \rightarrow \left\{\theta_{1\mathrm{true}},\ldots,\theta_{H\mathrm{true}}\right\}$ is always satisfied.

\subsection{Kolmogorov-Smirnov Parameter estimation}\label{subsect.KSparamestimate}

As discussed in section \ref{subsect:noisexcsmodel}, the KS estimator can be used as an alternative to a maximum likelihood procedure for parameter estimation.

Thanks to the statistical properties of $R_{\text{WOSA}}\left(f_k;\theta\right)$, the closer $\theta$ is to $\theta_{\mathrm{true}}$, the better the distribution of $R_{\text{WOSA}}\left(f_k;\theta\right)$
is described by equation (\ref{eqn.probdistwosa}) with $\delta = 1/N_s$ and $h = N_s$. 
Therefore, the Kolmogorov-Smirnov distance
parameter provides an effective goodness-of-fit estimator:
\begin{equation}
	d_K\left(\theta\right) = \max\left|F_R\left(x;\theta\right)-\mathcal{P}\left(N_s,x N_s\right)\right|.
\end{equation}
$F_R\left(\alpha\right)$ is the ECDF for the current  $R_{\text{WOSA}}\left(f_k;\theta\right)$ estimate, $\mathcal{P}\left(N_s,x N_s\right)$ is the limiting distribution for $\theta \rightarrow \theta_{\mathrm{true}}$. 

KS estimation for $\theta$ is obtained by the minimization of  $d_K\left(\theta\right)$
with respect to $\theta$. A confidence range for the parameter estimation can
be readily defined from the non-rejection region of the KS-test at a given significance
level. In practice, having defined a significance level, the corresponding
critical values of the KS statistic, $d_K\left(\alpha\right)$, can be calculated with standard equations
\cite{Miller1956} or Monte Carlo simulations
in the case of correlated data. The values of $\bar{\theta}$ for which $d_K\left(\bar{\theta}\right)
= d_K\left(\alpha\right)$ provide the boundary for the confidence range
at the given significance. The KS statistic can also be successfully applied to multi-parameter
identification, since the convergence of $F_R\left(\alpha\right)$ to $\mathcal{P}\left(N_s,x N_s\right)$ is always verified when $\left\{\theta_1,\ldots,\theta_H\right\} \rightarrow \left\{\theta_{1\mathrm{true}},\ldots,\theta_{H\mathrm{true}}\right\}$.


\section{Analysis of non-stationary noise}\label{sect.nonstatnoise}

The implementation of noise analysis procedures was so far discussed in the context
of stationary or {\it pseudo-stationary} noise \footnote{The term {\it pseudo-stationary}
indicates a data series which is affected by slight non-stationarity
of the kind that can be removed with mean subtraction or standard polynomial fit
trend removal.}. In the case of truly non-stationary noise the spectral content
of a time series is investigated by time-frequency analysis techniques which include
the spectrogram and the wavelet transform. 
The spectrogram is estimated by the square modulus of the
short-time Fourier transform of the data \cite{Oppenheim1999}. It provides a direct
extension of the PSD formalism to non-stationary time series. Given a data series
of $N$ samples, a fraction of length $Q < N$ is windowed and the Fourier transform
is applied. Then the data window is time shifted and the process is repeated.
The calculation of the spectrogram is based on data windowing and the application
of the Fourier transform, therefore the considerations noted in the previous sections for stationary
noise can be applied directly to the spectrogram analysis of non-stationary noise.

Since the short-time Fourier transform has the same resolution across the time-frequency plane, it is often preferable to resort to the wavelet transform.
Wavelet transform is a decomposition of the time series over time-frequency elements that are obtained by scaling and translating a mother function $\psi \in \mathbf{L^2}\left(\mathbb{R}\right)$:

\begin{equation}
 \psi_{u,s} = \frac{1}{\sqrt{s}} \psi \left(\frac{t-u}{s}\right).
\end{equation}

The function $\psi\left(t\right)$ has zero average and the wavelet elements $\psi_{u,s}$ are normalized to 1.
The wavelet transform of a function $f(t)$ is then defined as:

\begin{equation}
 Wf(u,s) = \int_{-\infty}^{\infty} f(t)\frac{1}{\sqrt{s}} \psi^* \left(\frac{t-u}{s}\right) dt.
\end{equation}

In the discrete case, the result of the wavelet transform on a time series is an array of coefficients $w_{u,s}$ where $u$ is the time index and $s$ is the scale index which is associated to a given frequency band \cite{MallatWavelet}.
In the case of uncorrelated Gaussian noise, the distribution of the coefficients $w_{u,s}$ is still Gaussian with a certain amount of correlation introduced by the convolution-like transform operation \cite{MallatWavelet}.
In such a favorable situation the extension of the method described in the stationary case it appears straightforward.
In particular Kolmogorov-Smirnov procedures are excellent candidates since their robustness to correlation and the possibility of extending KS distance definition to a two-dimensional space \cite{Peacock1983,Fasano1987}.


\section{Application to LPF synthetic data}\label{sect.testofperformance}

In this section the different procedures for excess noise detection and parameter estimation are applied to synthetic LPF data. This provides not only an interesting framework for testing their accuracy and precision, but it also helps clarifying the role of correlations among spectral data.

\subsection{Synthetic data and noise projection}

LPF provides two output channels along the principal measurement axis which are sensing the displacement of the SC relative to the free falling TM and the relative displacement between the TMs.

From the knowledge of the displacement signals and a linear model for the system dynamics, an effective force-per-unit-mass, $a_{\rm eff}$, acting on the TMs can be extracted. $a_{\rm eff}$ is the combination of the `true' force-per-unit-mass acting on TMs and a projected interferometer readout noise. Details of the calculation are reported in appendix \ref{appendix.conv2acc}. 

\begin{figure}
\includegraphics[width=0.45\textwidth]{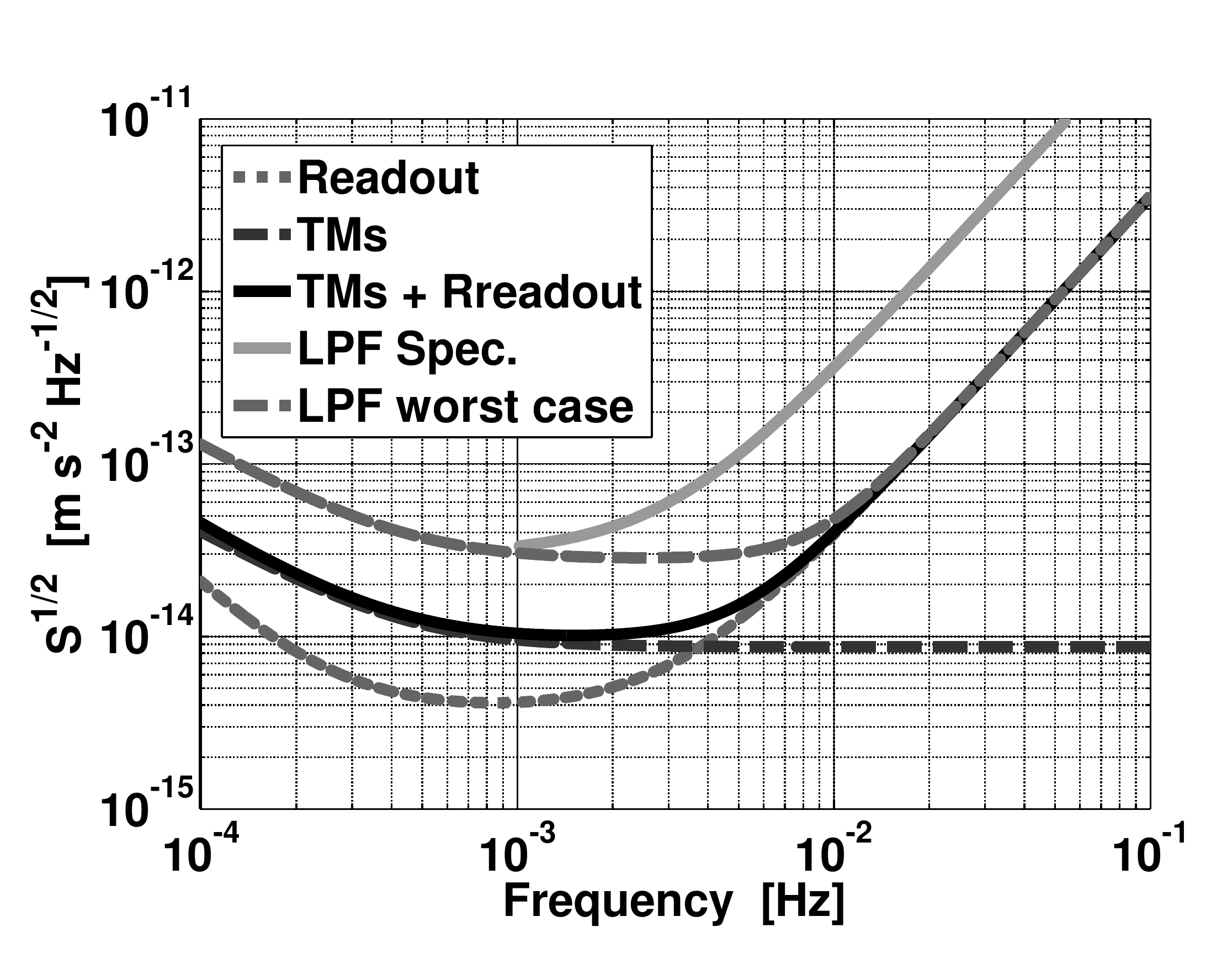}
\caption{Projection of the spectrum of $a_{\rm eff}$. `Readout' is the projection of the readout noise to $a_{\rm eff}$. `TMs' is the projection of the force noise on the test masses  to $a_{\rm eff}$. `TMs + Readout' is the complete noise projection for $a_{\rm eff}$, it represents the baseline noise level assumed in the present paper. `LPF worst case' refers to a worst case scenario for $a_{\rm eff}$ and `LPF Spec.' corresponds to the mission specifications.}
 \label{fig:fig4}
\end{figure}

Following the same scheme, a given prediction for the different input noise sources can be projected to a prediction for the power spectrum of $a_{\rm eff}$, as reported in figure \ref{fig:fig4}. Here the model for the spectrum of force noise acting on TMs and the model for the spectrum of readout noise are projected into a model for the power spectrum of $a_{\rm eff}$  (TMs + Readout in the figure notation), which represents this paper's baseline for the spectrum of $a_{\rm eff}$.   

In figure \ref{fig:fig4} we also report the project specifications for LPF and the expected noise spectrum for $a_{\rm eff}$ in a worst case scenario. In our baseline we assumed a reduced force noise on the TMs compared to the worst case but choose to keep the worst case for the readout noise. This was done in order to represent one of the possible scenarios (not the best one) that can be experienced during the mission.

The model assumed for the force noise on the TMs is characterized by a low frequency $1/f^2$ behavior and a flat part for $f > 1$ mHz. The model can be written as $S_{\rm TM}\left(f\right) = \theta S^0_{\rm TM}\left(f\right)$. $\theta$ is an adjustable parameter which assumes values $\theta = 1$ for the worst case scenario and $\theta = 0.1$ for our baseline model. 
It is worth noting that $S_{\rm TM}\left(f\right)$ is projected (together with the readout noise model) through LPF dynamics in order to obtain the expected noise spectrum for $a_{\rm eff}$ which we indicate with $S_a\left(f\right)$. $S_a\left(f\right)$ with $\theta = 0.1$ corresponds to `TMs + Readout' in figure \ref{fig:fig4}, $S_a\left(f\right)$ with $\theta = 1$ corresponds, instead, to the `LPF worst case'.

\subsection{Excess noise detection}\label{subsect.test.excessnoise}

A change of the noise level on $S_{\rm TM}\left(f\right)$ ($\theta \neq 0.1$) produces a variation of the energy content of $a_{\rm eff}$. Such variation, which may be `improperly' identified as excess noise, can be detected with the procedures defined in section \ref{sect.noisexcess}. In particular we tested the detection of excess noise between two data series and between a data series and a reference model.
Synthetic data were produced according to the following procedure: 
\begin{enumerate} 

\item Different models for $S_a\left(f\right)$ are produced changing $\theta$ around the reference value $\theta = 0.1$. Readout noise level is kept fixed.

\item Corresponding noise time series for $a_{\rm eff}$ are generated using the procedure reported in \cite{FerraioliPRD2010}. 
The time series are $24$ h long and have a sampling frequency of $1$ Hz.

\item Sample spectra are calculated for each series with the WOSA algorithm. We chose a Blackman-Harris data window, $50 \%$ segment overlap and number of segments averages $N_s = 4$.

\item The analysis is restricted to the frequency interval $\left[0.1, 10\right]$ mHz since, as can be seen from figure \ref{fig:fig4}, it represents the region in which the force noise on the TMs dominates $S_a\left(f\right)$.

\end{enumerate}

Spectral data are tested for excess noise. We used the KS algorithm (equation (\ref{eqn.dKdefemp})) in the case of the test of two data series. The data series for $\theta \neq 0.1$ are compared against the reference series with $\theta = 0.1$. In the case of the test of a data series against a model, both the KS algorithm (equation (\ref{eqn.dKdef})) and the IR algorithm (equation (\ref{eqn:IRdef})) are used. The reference model is the projected $S_a\left(f\right)$ for $\theta = 0.1$. The results of the tests are summarized in table \ref{tbl.testresults}. KS critical values $d_K\left(\alpha\right)$ and IR confidence intervals are calculated for a significance level $\alpha = 0.05$ which corresponds to a $95 \%$ confidence.

Each value of $\theta$ corresponds to a value of the in-band energy content $E\left(\theta\right)$ of $S_a\left(f\right)$ in the analyzed frequency band. We report in table \ref{tbl.testresults} the relative change in energy $\Delta E / E$ corresponding to a relative change in $\theta$. The quantity $\Delta E / E$ plays the same role of the parameter $\gamma$ in figures \ref{fig:fig1} and \ref{fig:fig3}, even though the two quantities are not completely comparable since $\gamma$ in figure \ref{fig:fig1} assumes independence of the data. Spectral data are correlated among different frequency values because of two effects \cite{Percival, Thomson1977}:
\begin{inparaenum}[\itshape  i\upshape)]
\item Data windowing which corresponds to a convolution in the frequency domain of the window function with the sample spectrum;
\item WOSA overlapped segment averaging.
\end{inparaenum}
The first effect is unavoidable, the second effect, instead, can be attenuated by a proper choice of the segment overlap. It can be demonstrated 
\footnote{If $N_s$ is the number of averaging segments, $N_f$ the length of each segment
and $h$ the shift factor, which is indicating the number of data points two consecutive
data segments are shifted by, then the correction to the variance
of the averaged process is proportional to \cite{Percival} $\Gamma\left(h\right) = \sum_{j=1}^{N_s-1}\sum_{t=1}^{N_f} h_t h_{t}^{j h}$.
Here $h_{t}^{j h}$ are window elements shifted by a factor $j h$. With our choice of $N_s = 4$, a segment overlap of $50 \%$ and a total data length of $24$ h, we find $\Gamma\left(h\right) = 2.7 \times 10^{-4}$. } 
that for a Blackmann-Harris window the effect is negligible with $50 \%$ overlap.

\begingroup
\squeezetable
\begin{table*}[h]
  \caption{Detection of noise differences in the frequency band [0.1, 10] mHz. The symbol $\surd$ indicates compatibility between tested objects. The symbol $\times$ is instead used for indicating test rejection. $\theta$ is an adjustable parameter which assumes value $\theta = 0.1$ for our baseline model. Different values of $\theta$ correspond to different values of the in-band energy content $E\left(\theta\right)$ of $S_a\left(f\right)$. We reported here relative values with respect to reference. Details on $d_K^{\rm eff}\left(\alpha\right)$, $d_K^{\rm MC}\left(\alpha\right)$ and MC confidence interval for IR estimator can be found in appendix \ref{appendix.KStest} and \ref{appendix.IRtest} respectively.
\label{tbl.testresults}}
\begin{ruledtabular}
\begin{tabular}{ c c | c  c  c | c  c  c  c | c  c  c }
\hline
\hline
 \multicolumn{2}{c}{} & \multicolumn{3}{c}{KS vs. data} & \multicolumn{4}{c}{KS vs. model} & \multicolumn{3}{c}{IR} \\
 \hline
  &  &   & $d_K\left(\alpha\right)$ & $d_K^{\rm MC}\left(\alpha\right)$ &  & $d_K\left(\alpha\right)$ & $d_K^{\rm eff}\left(\alpha\right)$ & $d_K^{\rm MC}\left(\alpha\right)$ &  &	Conf. Int. & MC Conf. Int. \\
 $\mathbf{\frac{\Delta E}{E}}$ & $\mathbf{\frac{\Delta\theta}{\theta}}$ & $d_K$ & $0.1030$ & $0.1006$ & $d_K$ & $0.0730$ & $0.0982$ & $0.0969$ & $\sum_k{R_{\text{WOSA}}\left(f_k\right)}$ &	$\left[323.14, 359.33\right]$ & $\left[311.33, 372.55\right]$ \\
\hline
$-0.14$ & $	-0.7	 $ & $	0.2352	 $ & $	\times	 $ & $	\times	 $ & $	0.2547	 $ & $	\times	 $ & $	\times	 $ & $	\times	 $ & $	250.74	 $ & $	\times	 $ & $	\times	$ \\
$-0.08$ & $	-0.4	 $ & $	0.1424	 $ & $	\times	 $ & $	\times	 $ & $	0.1103	 $ & $	\times	 $ & $	\times	 $ & $	\times	 $ & $	303.72	 $ & $	\times	 $ & $	\times	$ \\
$-0.06$ & $	-0.3	 $ & $	0.1806	 $ & $	\times	 $ & $	\times	 $ & $	0.1747	 $ & $	\times	 $ & $	\times	 $ & $	\times	 $ & $	274.54	 $ & $	\times	 $ & $	\times	$ \\
$-0.05$ & $	-0.25	 $ & $	0.1740	 $ & $	\times	 $ & $	\times	 $ & $	0.1685	 $ & $	\times	 $ & $	\times	 $ & $	\times	 $ & $	294.42	 $ & $	\times	 $ & $	\times	$ \\
$-0.04$ & $	-0.2	 $ & $	0.0887	 $ & $	\surd	 $ & $	\surd	 $ & $	0.0696	 $ & $	\surd	 $ & $	\surd	 $ & $	\surd	 $ & $	317.10	 $ & $	\times	 $ & $	\surd	$ \\
$-0.03$ & $	-0.15	 $ & $	0.0906	 $ & $	\surd	 $ & $	\surd	 $ & $	0.0433	 $ & $	\surd	 $ & $	\surd	 $ & $	\surd	 $ & $	324.87	 $ & $	\surd	 $ & $	\surd	$ \\
$-0.01$ & $	-0.05	 $ & $	0.0771	 $ & $	\surd	 $ & $	\surd	 $ & $	0.0302	 $ & $	\surd	 $ & $	\surd	 $ & $	\surd	 $ & $	339.47	 $ & $	\surd	 $ & $	\surd	$ \\
$0.01$ & $	0.05	 $ & $	0.0551	 $ & $	\surd	 $ & $	\surd	 $ & $	0.0539	 $ & $	\surd	 $ & $	\surd	 $ & $	\surd	 $ & $	352.26	 $ & $	\surd	 $ & $	\surd	$ \\
$0.03$ & $	0.15	 $ & $	0.0629	 $ & $	\surd	 $ & $	\surd	 $ & $	0.0370	 $ & $	\surd	 $ & $	\surd	 $ & $	\surd	 $ & $	345.38	 $ & $	\surd	 $ & $	\surd	$ \\
$0.04$ & $	0.2	 $ & $	0.0603	 $ & $	\surd	 $ & $	\surd	 $ & $	0.0635	 $ & $	\surd	 $ & $	\surd	 $ & $	\surd	 $ & $	357.14	 $ & $	\surd	 $ & $	\surd	$ \\
$0.05$ & $	0.25	 $ & $	0.0519	 $ & $	\surd	 $ & $	\surd	 $ & $	0.0818	 $ & $	\times	 $ & $	\surd	 $ & $	\surd	 $ & $	369.05	 $ & $	\times	 $ & $	\surd	$ \\
$0.06$ & $	0.3	 $ & $	0.0857	 $ & $	\surd	 $ & $	\surd	 $ & $	0.1253	 $ & $	\times	 $ & $	\times	 $ & $	\times	 $ & $	388.01	 $ & $	\times	 $ & $	\times	$ \\
$0.08$ & $	0.4	 $ & $	0.1103	 $ & $	\times	 $ & $	\times	 $ & $	0.1137	 $ & $	\times	 $ & $	\times	 $ & $	\times	 $ & $	402.14	 $ & $	\times	 $ & $	\times	$ \\
$0.14$ & $	0.7	 $ & $	0.1520	 $ & $	\times	 $ & $	\times	 $ & $	0.2423	 $ & $	\times	 $ & $	\times	 $ & $	\times	 $ & $	459.64	 $ & $	\times	 $ & $	\times	$ \\
\hline
\end{tabular}
\end{ruledtabular}
\end{table*}
\endgroup

Since the standard statistics for the estimators ($d_K\left(\alpha\right)$ for KS and confidence interval for IR) are calculated in section \ref{sect.noisexcess} with the assumption of data independence, the standard critical values, $d_K\left(\alpha\right)$, for the KS estimators and the confidence intervals for the IR estimator can be applied only if the correlations among data are negligible. If this is not the case, the effective statistic of the estimators can be numerically calculated with Monte Carlo simulations. The corresponding results of a Monte Carlo simulation with $N_{\rm MC} = 5000$ realizations of the reference data series are indicated in table \ref{tbl.testresults} with the suffix MC. The symbol $\surd$ is used to indicate that the spectral data for the corresponding $\Delta E/ E$ is compatible with the reference. On the contrary the symbol $\times$ indicates a rejection.

Observing the test results reported in columns $7$ and $9$ for the KS algorithm and the results in columns $11$ and $12$ for the IR algorithm it is readily seen that correlations among data play a role. In the case of the KS test, the comparison of the data with $d_K\left(\alpha\right)$ determines a rejection for $\Delta E/ E = 0.05$. On the other hand, the same value is accepted when Monte Carlo result $d_K^{\rm MC}\left(\alpha\right)$ is used. In the case of the IR estimator, the comparison with the standard confidence interval leads to the rejection in correspondence with $\Delta E / E = 0.05$ and $\Delta E / E = -0.04$. Such values are instead considered compatible by the Monte Carlo confidence interval. These results provide us with clear information that the presence of correlations among data has affected the tests statistics and therefore the standard equations, assuming independence among data, are not usable in this situation. 

Looking at the results for the KS test between two data series, we discover that $d_K\left(\alpha\right)$ and $d_K^{\rm MC}\left(\alpha\right)$ are practically equal. In fact the results for the two corresponding columns of table \ref{tbl.testresults} (columns $4$ and $5$) are in perfect agreement. This is the practical result of one of the most interesting properties of the KS test. The test is based on equation (\ref{eqn.dKdefemp}). It states that the two empirical cumulative distributions under test have the same limiting CDF. Since the sample spectra in our test are calculated following the same WOSA procedure, the degree of correlation among different frequency points is the same, therefore the test statistic is not spoiled. 

Comparing $d_K\left(\alpha\right)$ and $d_K^{\rm MC}\left(\alpha\right)$ of columns $7$ and $9$ we see that the effect of correlation is to increase the maximum expected spread between ECDF and limiting CDF. Therefore the effect of data correlation is to distort the expected statistic for $d_K$. The `distortion' of $d_K$ statistic can be taken into account if an effective value for the parameter $K$ is introduced. In the case of the comparison of the ECDF for correlated data against a theoretical CDF the application of the standard values for $d_K$, where $K = N_f$, $N_f$ being the number of data elements, leads to a statistically unfair test. We then discovered that test fairness can be recovered if an effective value for $K$ is used rather than the standard $K = N_f$. In particular, for spectral data produced with the WOSA method, Blackmann-Harris window, $N_s = 4$ averages on $50 \%$ overlapped segments, we obtained $K_{\rm eff} = \beta N_f$ with $\beta = 0.55$ for a significance level $\alpha = 0.05$. It is worth noting that the value of $\beta$ is independent from the number of data points considered, and from the spectral shape, provided that the different shapes have reasonably comparable smoothness on a frequency interval comparable with the width of the first lobe of the data window. As an example, the value of $\beta = 0.55$ is valid for LPF-like data and for white-noise equivalently. The requirement on the smoothness of the spectrum is connected to the expression of the correlations introduced by data windowing. It can be demonstrated that if the spectrum can be assumed constant in a frequency range of the order of the width of the first lobe of data window \cite{Percival} then the correlations are independent from the particular shape of the spectrum and are determined only by the window function. For such class of spectra we expect the same value for $\beta$ once the required significance level is fixed.

\subsection{Noise model identification}\label{subsect.test.noisemodelident}

\begin{figure*}
\includegraphics[width=0.80\textwidth]{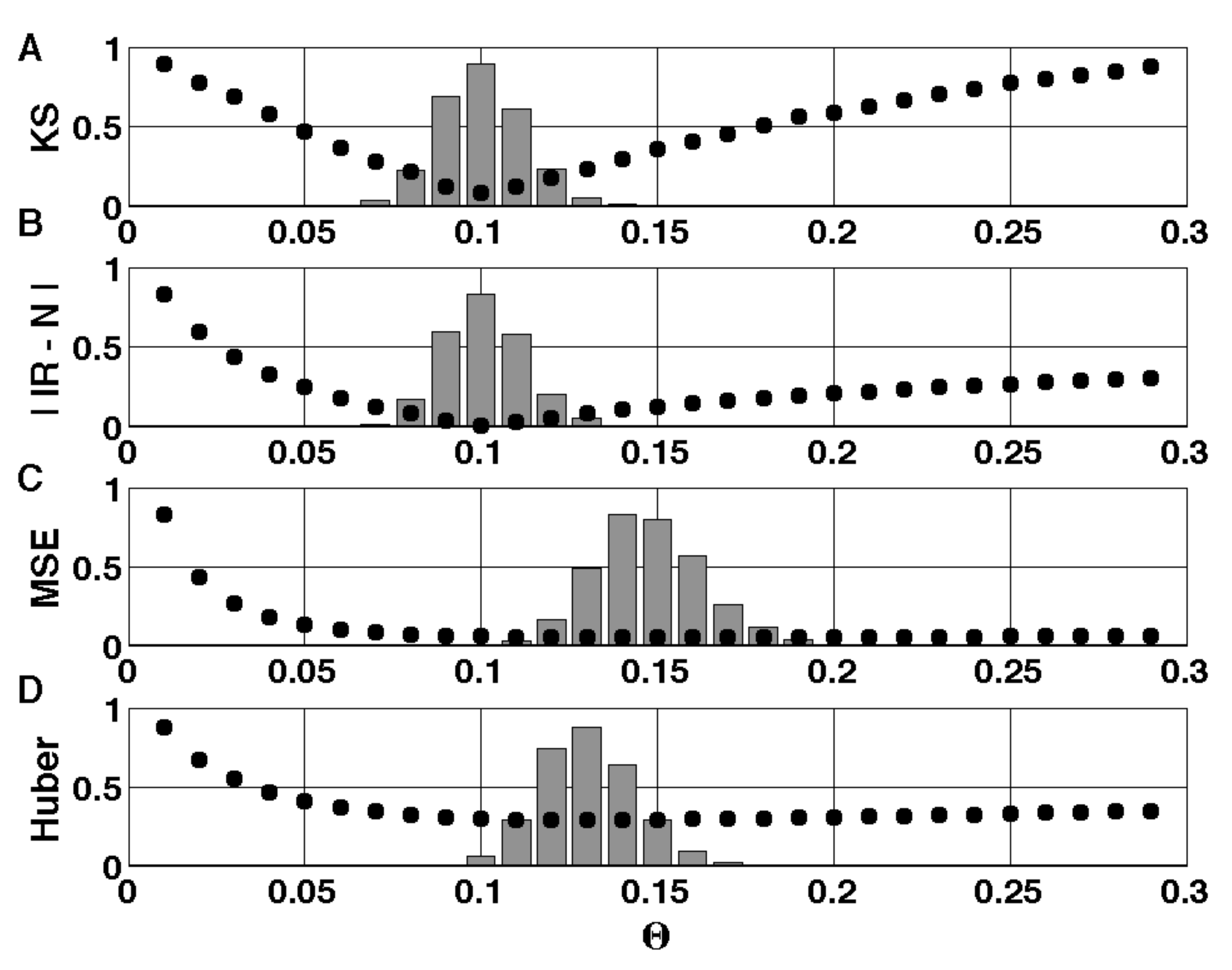}
\caption{KS, IR, MSE and Huber's norm accuracy and precision in the estimation of the force-per-unit-of-mass noise parameter $\theta$. Corresponding histograms report the result of $N_{\rm MC} = 5000$ MC realizations of the parameter search on a grid of template models. We also show (black dots) the evolution along the model grid of the four estimators for a particular set of data taken from the $N_{\rm MC}$ available.}
 \label{fig:fig5}
\end{figure*}

The problem of noise parameter identification is strictly connected to the problem of excess noise detection by a comparison of a data series with a reference model. While in excess noise detection different data series are compared with a given reference, in parameter estimation different realizations of a parametric model are compared with a dataset in order to find the best fit. Due to this, the same algorithms (i.e. KS and IR) can be applied to the solution of the two problems.
Precision and accuracy in the estimation of the parameter, $\theta$, controlling the excess force noise on the TMs is tested with a Monte Carlo simulation over $N_{\rm MC} = 5 \times 10^3$ realizations of the same process. The data series reproduce $a_{\rm eff}$ corresponding to the reference value $\theta = 0.1$. Sample spectra are calculated with the procedure described above and the analysis is restricted to the frequency range $\left[0.1, 10\right]$ mHz.
For each realization, data are compared with a bank of models obtained by the projection of TMs force noise $S_{\rm TM}\left(f;\theta\right) = \theta S^0_{\rm TM}\left(f\right)$ and readout noise into $S_a\left(f;\theta\right)$ for different values of $\theta$ around the reference value.
The KS and IR estimators are calculated for each element of the model bank, in particular it is simpler to analyze the results in terms of $\tilde{\mathrm{IR}} = \left|\mathrm{IR} - N_f\right|$. Both KS and $\tilde{\mathrm{IR}}$ are expected to have a minimum corresponding to the best estimate for the parameter $\theta$.
The two methods proposed here are compared with the performance of a classical weighted least-squares method, which works by minimizing the mean squared error $\mathrm{MSE} = \sum_{f_k}{\left(\left(P_{WOSA}\left(f_k\right)-S_a\left(f_k;\theta\right)\right) / S_a\left(f_k;\theta\right)\right)^2 }$, and a Huber's norm estimator (details are reported in appendix \ref{appendix.huber}).
The results of the analysis are reported in figure \ref{fig:fig5}, where the histograms of the best fit vales over $N_{\rm MC}$ realizations are reported for the four procedures.
In the same figure we also show the evolution along the model grid of the four estimators for a particular set of data from the available $N_{\rm MC}$. 

The distributions for the best fit parameter are reasonably symmetric for all the estimators; mean values and sample standard deviations are respectively $\theta_{KS} = 0.100$, $\sigma_{KS} = 0.012$, $\theta_{IR} = 0.100$, $\sigma_{IR} = 0.012$, $\theta_{MSE} = 0.148$, $\sigma_{MSE} = 0.016$, $\theta_{Huber} = 0.130$ and $\sigma_{Huber} = 0.014$.
From the analysis of the Monte Carlo results, it is readily seen that both the KS and the $\tilde{\mathrm{IR}}$ estimators provide equivalently precise and accurate results. On the other hand, the MSE algorithm provides a poor estimation, both from the accuracy and from the precision point of view. The best estimation for the parameter is $\theta_{MSE} = 0.148$, which is strongly biased with respect to the reference value of $\theta = 0.1$. Also, the distribution of the parameter values is wider than those obtained from the KS and $\tilde{\mathrm{IR}}$ estimators. Huber's norm estimator, with the chosen parameter $c = 0.05$ \ref{appendix.huber}, performs better than MSE but the result $\theta_{Huber} = 0.130$ is still far from the true value.

It is worth discussing the effect of correlations among spectral data on the estimation of the parameter $\theta$. As can be observed from the results of the MC simulation, the accuracy of the estimators is not affected by correlations; the results for KS and $\tilde{\mathrm{IR}}$ estimators are practically indistinguishable. Some problems with $\tilde{\mathrm{IR}}$ can arise from data correlations when the confidence interval for a single estimation is required. As discussed in the previous paragraph, correlations modify the statistic of the IR estimator making it impossible to easily calculate confidence intervals from the standard equations. The statistic of the KS estimator is affected by correlations too, but we have demonstrated that accurate critical values $d_K\left(\alpha\right)$ can be recovered if an effective value for $K$ is used, $K_{\rm eff} = \beta N_f$. $\beta$ is a shape parameter depending only on the correlations in the spectral data and on the required significance level. It can be calculated for a specific spectrum (e.g., white noise) and effectively extended to a wide family of spectral shapes. Using the value $\beta = 0.65$ reported in appendix \ref{appendix.KStest}, we obtain $d_K^{\rm eff}\left(\alpha\right) \approx 0.064$ for $\alpha = 1 - 0.68 \%$. The intersection of such a value with the curve of $d_K$ as a function of $\theta$ reported as black dots in figure \ref{fig:fig5}A provides a $68 \%$ confidence interval for the single estimation. In this specific case such an interval is $\theta \in \left[0.086, 0.114\right]$.


\section{Conclusions}\label{conclusions}

The problem of excess noise detection and noise parameter estimation for non-Gaussian data is analyzed in the framework of the LISA Pathfinder mission.
Excess noise detection can be approached in two ways. In one way, the noise content of a data series is compared with a reference data series, in the other way the noise content of the data is compared with a reference model.
In the first case, simple estimators like the total energy content in a data series are not suitable for formulating quantitative statements on a solid statistical basis. As an alternative a Kolmogorov-Smirnov (KS) estimator is proposed and successfully applied to LPF synthetic data. The KS estimator has the advantage of being independent of the statistical properties of the data under test. It is demonstrated in the paper that such a convenient property makes the estimator robust to correlations among spectral data.
Two different estimators are investigated for the problem of comparing a data series to a model for excess noise detection. One estimator (IR) is based on the integral of the normalized WOSA spectrum, the other is a KS estimator for the comparison of an empirical cumulative distribution function with a limiting theoretical function.
Despite the fact that IR estimator proves to have better sensitivity on independent data, the versatility of the KS estimator is highly advantageous on correlated data.
Since the statistics of the estimators are distorted by the presence of correlations among test data, the standard and simple procedures for the determination of the confidence intervals, based on the inversion of the limiting distribution function, provide inaccurate results.
While, in the case of the IR estimator, the problem can be overcome only with dedicated Monte Carlo simulations, we demonstrated that the introduction of a shape parameter allows us to use standard equations to calculate proper boundaries for KS confidence intervals. The shape parameter depends on the required significance level and on the data correlations. It does not depend on the number of data considered for the test. Since correlations among spectral data are mainly introduced by the windowing process, the shape parameter is fixed for a wide family of spectral shapes. For example, synthetic LPF data share the same shape parameter with white noise.

Closely related to noise excess detection, the problem of noise parameter identification is analyzed with a maximum likelihood approach which, in the particular case of linear dependence on a single parameter, provides an algorithm analogous to the IR estimator used for excess noise detection. A KS algorithm was proposed as an alternative to the IR algorithm and the accuracy and precision of both were tested with a Monte Carlo simulation on LPF synthetic data.
Both the IR and KS estimators were demonstrated to give equivalently good results, even though the capability of the KS to handle data correlation is a clear advantage for the definition of a confidence interval for the estimated noise parameter. 

Data analysis procedures introduced in this paper are easily extended to the vast
context of time-frequency analysis of non-stationary noise. KS algorithm can be applied effectively both to spectrogram and wavelets coefficients provided that correlations among data are taken into account. KS algorithm has the advantage that can be generalized to two dimensions thus allowing to extend the analysis to the time-frequency plane.


\appendix

\section{Statistic of the sample spectrum}\label{sec.spectrumstat}

In the case of a discrete, real-valued stationary process $\left\{x_h\right\}$, the continuous spectral density function is defined as:
\begin{equation}
	S\left(f\right) = T \sum_{h=-\infty}^{\infty}{s_h e^{- \imath 2 \pi f h T}}.
\end{equation}
Here $s_h$ is the autocovariance sequence of the process $\left\{x_h\right\}$, $T$ is the sampling period and $f$ is the frequency expressed in Hz. $f$ is defined in the range $\left|f\right| \leq f_{Nq} \equiv 1/2T$ and $f_{Nq}$ is known as the Nyquist frequency.
In the case of a finite representation $x_0,\ldots,x_{N-1}$ of the discrete process $\left\{x_h\right\}$, the approximation to the spectral density function is provided by the sample spectrum
\begin{equation}\label{eqn.sampspect}
	\tilde{P}\left(f\right) = \frac{T}{N}\left|\sum_{h=0}^{N-1}{x_h e^{- \imath 2 \pi f h T}}\right|^2.
\end{equation}
$f$ in this case is also defined on the interval $\left[-f_{Nq},f_{Nq}\right]$. If the sample spectrum is calculated on the grid of Fourier frequencies ($f_k = k/\left(N T\right)$, $\left|k\right| \leq N/2$) then it corresponds to the squared modulus of the discrete Fourier transform of the data sequence $x_0,\ldots,x_{N-1}$. In practical applications only the positive frequency part of the spectrum is considered, and the one sided sample spectrum is defined as $P\left(f_k\right) = 2 \tilde{P}\left(f_k\right)$ with $k = 0, 1, \ldots, N/2$. In the rest of the paper the one sided sample spectrum will be simply named the sample spectrum.

If the data series $x_0,\ldots,x_{N-1}$ is Gaussian distributed and the elements $x_j$ are independent then the Fourier transform produces a complex series $X\left(f_k\right)$ whose elements are approximately independent and their real and imaginary parts are Gaussian distributed. The term $\left|X\left(f_k\right)\right|^2 = \left|\Re\left[X\left(f_k\right)\right]\right|^2 + \left|\Im\left[X\left(f_k\right)\right]\right|^2$ is then the sum of two independent variables distributed as a $\chi^2_{\nu}$ where $\nu$ is the number of degrees-of-freedom of the distribution ($\nu = 1$ in this case).
If the correlations among the elements of the data series $x_0,\ldots,x_{N-1}$ are non-vanishing, the statistical properties of the sample spectrum, in the simplifying assumption of independent $P\left(f_k\right)$ elements, can be calculated from the case of a $\chi^2_{\nu}$ distributed variable $y$ multiplied by a constant $z = \lambda y$.
The characteristic function for $z$ is
\begin{equation}\label{eqn.sampspcritical}
	\phi_z\left(t\right) = E\left[e^{\imath t \lambda y}\right] = \left(1 - 2 \imath t \lambda \right)^{-\frac{\nu}{2}}.
\end{equation}
Here $E\left[\right]$ indicates the expected value.
The inverse Fourier transform of $\phi_z\left(t\right)$ provides the probability density function for $z$:
\begin{equation}\label{eqn.sampspprobdist}
	\mathcal{F}^{-1}\left[\phi_z\left(t\right)\right] = \frac{e^{-\frac{z}{2\lambda}} z^{\left(\frac{\nu}{2}-1\right)}}{\left(2\lambda\right)^{\frac{\nu}{2}}\Gamma\left(\frac{\nu}{2}\right)}.
\end{equation}

This is a gamma distribution, $f\left(z;k,\theta\right)$, with $k = \nu/2$ and $\theta = 2\lambda$. $\Gamma\left(k\right)$ is the gamma function. In the case of the sample spectrum at a given frequency, $z = P\left(f_k\right)$ and $\lambda = E\left[P\left(f_k\right)\right]/\nu \approx S\left(f\right)/\nu$
\footnote{It is easy to verify that, in the case of $z = P\left(f_k\right)$ and $\lambda = E\left[P\left(f_k\right)\right]/\nu$, the probability density function $f\left(z;\nu,\lambda\right) = \frac{e^{-\frac{z}{2\lambda}} z^{\left(\frac{\nu}{2}-1\right)}}{\left(2\lambda\right)^{\frac{\nu}{2}}\Gamma\left(\frac{\nu}{2}\right)}$ is correctly represents the statistic of the sample spectrum. $E\left[z\right] = \frac{\nu}{2}2\lambda = E\left[P\left(f_k\right)\right]$ as expected for the spectrum \cite{Percival}. ${\rm var}\left[z\right] = \frac{\nu}{2} \left(2 \lambda\right)^2 = \frac{2}{\nu} E\left[P\left(f_k\right)\right] = E\left[P\left(f_k\right)\right]$ since $\nu = 2$ in the present case.}. 
It is useful, for the statistical analysis of the spectrum, to introduce the normalized sample spectrum
\begin{equation}
	R\left(f_k\right) = \nu \frac{P\left(f_k\right)}{S\left(f\right)},
\end{equation}
which, at each frequency $f_k$, is distributed as $\chi^2_{\nu}$.

The sample spectrum in equation (\ref{eqn.sampspect}) can also be seen as a special case of the power spectral density, $S\left(f\right)$, when the infinite data series $\left\{x_h\right\}$ is chopped by a square data window of length $N$. This operation introduces a considerable amount of spectral leakage because of the convolution with the frequency response of the square data window \cite{Percival, Harris1978}. Therefore it is common practice to multiply the time series $x_0,\ldots,x_{N-1}$ with a more performant data window, which increases the accuracy of the sample spectrum in the case of processes with a high dynamic range. The application of a data window introduces correlations among different elements of the sample spectrum. Such correlations affect the statistics of the spectrum, resulting in a change in the probability distribution of the sample spectrum. An analytical treatment of the spectrum statistics under such conditions is cumbersome, and it is easier to numerically evaluate (with a Monte Carlo simulation) the statistics of the sample spectrum for the case of interest.

In order to improve the variance properties of the sample spectrum, Welch's overlapped segment averaging (WOSA) method is applied \cite{Percival}. The data series $x_0,\ldots,x_{N-1}$ is divided in overlapping windowed segments. The estimates of the sample spectrum of each segment are then averaged. The practice of averaging overlapping segments can modify the expected statistics of the spectrum since the data in different segments can be correlated. Then, even in the simplifying assumption of vanishing spectral correlations from windowing and overlapping, the averaging process changes the statistics of the estimated sample spectrum.
In the assumption of vanishing window and overlap correlations, the statistic of the WOSA spectrum corresponds to the average of $N_s$ gamma distributed variables
\begin{equation}
	P_{\text{WOSA}}\left(f_k\right) = \frac{1}{N_s}\sum_{j=1}^{N_s}{P_j\left(f_k\right)}.
\end{equation}

The critical function for the sum is $\phi_{\text{P}_{\text{WOSA}}}\left(t\right) = \prod_j \phi_{j}\left(t\right)$, where $\phi_j\left(t\right)$ is the critical function for $\text{P}_j/N_s$. Thanks to equation (\ref{eqn.sampspcritical})
\begin{equation}
	\phi_{\text{P}_{\text{WOSA}}}\left(t\right) = \left(1 - \frac{\imath t \theta}{N_s}\right)^{-k N_s},
\end{equation}
where $\theta = 2 S\left(f_k\right)/\nu$ and $k = \nu/2$. The inverse Fourier transform of $\phi_{\text{P}_{\text{WOSA}}}\left(t\right)$ provides the probability density function for the WOSA spectrum
\begin{equation}\label{eqn.probdistwosa}
	f_{\text{WOSA}}\left(z; h, \delta\right) = \frac{z^{\left(h-1\right)} e^{-\frac{z}{\delta}}}{\delta^h \Gamma\left(h\right)},
\end{equation}

with $\delta = S\left(f_k\right)/N_s$ and $h = N_s$. Again it is useful to define a normalized WOSA spectrum as
\begin{equation}\label{eqn.normwosa}
	R_{\text{WOSA}}\left(f_k\right) = \frac{P_{\text{WOSA}}\left(f_k\right)}{S\left(f_k\right)},
\end{equation}

which is gamma distributed (equation (\ref{eqn.probdistwosa})) with $\delta = 1/N_s$ and $h = N_s$.


\section{Kolmogorov - Smirnov Test} \label{appendix.KStest}

Kolmogorov - Smirnov is a well known test for distributions \cite{Kolmogorov1933,
Smirnov1939, Feller1948, Miller1956, Fisher1983, Wilk1968}. An empirical cumulative
distribution (ECDF) is tested against a continuous theoretical model or, alternatively,
two ECDFs are tested with the hypothesis that they share the same limiting cumulative distribution
function. Indicating with $f\left(x\right)$ the probability density function
associated with a given random process $X$, the corresponding cumulative distribution
function (CDF) is defined as:
\begin{equation}
	F\left(x\right) = Prob\left[X \leq x \right] = \int_{-\infty}^x f\left(u\right) \, du.
\end{equation}

Given a particular realization of the random process $X$:
\begin{equation}
	X_N = \left\{x_1, \ldots, x_N \right\},
\end{equation}
ECDF is written as $F_N\left(x\right) = k/N$ where $k$ is the number of observations
which is smaller or equal to $x$.

Given two data series $X_N$ and $Y_M$, with $N$ and $M$ not necessarily equal,
we can test if the two series are two particular realizations of the same
random variable by analysis of their ECDFs.
Under the hypothesis that the two data series comes from the same distribution function, Kolmogorov has demonstrated that the maximum distance between the two ECDFs has a limiting distribution which is independent from the statistical properties of the corresponding random variable.
If the test is performed against a theoretical distribution, the distance is defined as:
\begin{equation}
	d_K = max\left|F_N\left(x\right) - F\left(x\right)\right|.
\end{equation}

In such a case $K = N$.
In alternative, If the test is performed between two ECDFs, $K = \left(N M\right)/ \left(N + M\right)$ and:

\begin{equation}
	d_K = max\left|F_N\left(x\right) - F_M\left(x\right)\right|.
\end{equation}

The test is defined as follows:

\begin{enumerate}
  
\item In the case of the test on a single data series, the null hypothesis is that the
data are realizations of a random variable which is distributed according to
the given probability distribution. In the case of two data series, the null hypothesis
is that the two data series are realizations of the same random variable, which
means their ECDFs should tend to the same CDF. The test rejects or accepts the null hypothesis on the basis of the analysis of $d_{nm}$.

\item A significance level $\alpha$ is defined as the probability that
the test rejects the null hypothesis when it is indeed true.

\item The test can be formulated in terms of critical values. The critical value
for the test is the value of $d_K\left(\alpha\right)$ corresponding to the significance level. Then if $d_K > d_K\left(\alpha\right)$, the null hypothesis is rejected.

\end{enumerate}

KS theory was formulated for independent data sets and the available equations for critical
values are valid only if this condition is satisfied \cite{Miller1956}.
The test can be formulated also in the presence of data correlation but the distortion to $d_K$ statistic introduced should be taken into account. 
This is possible if an effective value for $K$ is introduced as $K_{\rm eff} = \beta K$, with $\beta$ a shape parameter depending only on the data correlations and the required significance level. Alternatively, realistic critical values can be calculated with dedicated Monte Carlo (MC) simulations \cite{Weiss1978}.
The advantage of the shape parameter $\beta$ is that it depends only on correlations and the required significance level. Therefore it can be determined for a specific spectrum (e.g., white noise) and shared among a wide family of spectral shapes. Once $\beta$ is known it can be used to calculate critical values for correlated data using standard equations reported in the literature \cite{Miller1956}.

Focusing on the particular problem, we performed a Monte Carlo estimation of $d_K\left(\alpha\right)$ for WOSA spectra representative of LPF. The number of frequency data considered is $N_f = 341$, corresponding $K$ values are $K = N_f$ in the case of the test against a theoretical distribution and $K = N_f/2$ in the case of the test between two ECDFs. The results are summarized in table \ref{tbl.kscritmc}. In the same table we report the values of $\beta$ that are required to obtain proper critical values from the standard equations in the case of the test against a theoretical distribution.

\begingroup
\squeezetable
\begin{table}[h]
\caption{Table of KS critical values for correlated spectral data. Critical values are calculated with a Monte Carlo simulation for different values of the significance level $\alpha$. Datasets used are representative of the data analyzed in the current paper. We reported the values for testing an ECDF against a theoretical CDF ($K = N_f$) and the values for testing two ECDFs for the same limiting CDF ($K = N_f/2$). We also reported the values of the shape parameter $beta$ that can be used to calculate proper critical values from standard equations in the case of correlated data. $\alpha$ refers to the significance level whereas $1-\alpha$ is the corresponding confidence level for the test. $N_f = 341$. \label{tbl.kscritmc}}
\begin{ruledtabular}
\begin{tabular}{ l l l l l}

\bf $\alpha$ &	\bf $1-\alpha$ &	\bf $K = N_f$ & $\beta$ &	\bf $K = N_f/2$ \\
\hline
$0.32$  &	$0.68$ & 	$0.0643$ & $0.65$ &	$0.0723$ \\
$0.10$ &	$0.90$ &	$0.0863$ & $0.58$ & $0.0910$ \\
$0.05$  &	$0.95$ &	$0.0969$ & $0.55$ & $0.1006$ \\
$0.01$ &	$0.99$ & 	$0.1214$ & $0.52$ &	$0.1191$ \\

\end{tabular}
\end{ruledtabular}
\end{table}
\endgroup


\section{IR Test} \label{appendix.IRtest}

The statistics of the IR excess noise estimator are numerically estimated with a Monte Carlo simulation on the spectral data used for the present paper. Data series are $24$ h long synthetic reconstructions of the force-per-unit-mass expected on the TMs. WOSA spectra are calculated with $N_s = 4$ averages on $50 \%$ overlapped segments which were windowed with the Blackman-Harris window. Results are reported in table \ref{tbl.IRconf}.

\begingroup
\squeezetable
\begin{table}[h]
\caption{Table of confidence bounds for the IR estimator on correlated spectral data. The values are calculated with a Monte Carlo simulation for different values of the significance level $\alpha$. Data sets used are representative of the data analyzed in the current paper. The number of frequency points is $N_f = 341$, corresponding to the number of available spectral data in the range $\left[0.1, 10\right]$ mHz, for a time series $24$ h long and number of averages for the WOSA estimator $N_s = 4$. \label{tbl.IRconf}}
\begin{ruledtabular}
\begin{tabular}{ l l l l}

\bf $\alpha$ &	\bf $1-\alpha$ &	\bf $x_{lw}$ &	\bf $x_{up}$ \\
\hline
$0.32$  &	$0.68$ & 	$325.88$  &	$357.41$ \\
$0.10$ &	$0.90$ &	$316.33$  &  $367.28$ \\
$0.05$  &	$0.95$ &	$311.33$  &  $372.55$ \\
$0.01$ &	$0.99$ & 	$300.48$  &	$380.69$ \\

\end{tabular}
\end{ruledtabular}
\end{table}
\endgroup


\section{Conversion of displacement noise}\label{appendix.conv2acc}

LPF can be considered as a three body controlled system composed of the two TMs and the spacecraft (SC). Its equations of motion along the measurement axis can be written as:
\begin{equation}\label{eqn:lpfeqmotion1}
	\begin{split}
		& m_1 \ddot{x_1} + m_1 \ddot{x_{sc}} + m_1 \omega_1^2 x_1 = f_1 \\
		& m_2 \ddot{x_2} + m_2 \ddot{x_{sc}} + m_2 \omega_2^2 x_2 = f_2 + f_{c2} \\
		& m_{sc} \ddot{x_{sc}} - m_1 \omega_1^2 x_1 - m_2 \omega_2^2 x_2 = f_{sc} - f_{c2} + f_{csc}\,. \\
	\end{split}
\end{equation}

Here:
\begin{itemize}
	\item $x_1$ and $x_2$ are TMs coordinates along the sensitive axis. They are relative coordinates in the SC reference frame.
	\item $x_{sc}$ is the absolute SC coordinate along the sensitive axis.
	\item $m_1$, $m_2$ and $m_{sc}$ are the masses of the two TMs and of the SC.
	\item $\omega _1^2$ and $\omega _2^2$ are the parasitic stiffnesses coupling the TMs and the SC. The TMs are coupled to the spacecraft through the parasitic stiffness thus producing an oscillator like equation of motion. The spacecraft at the same time experiences reaction forces given by $- m_1 \omega _1^2 x_1$ and $ - m_2 \omega _2^2 x_2$.
	\item $f_1$, $f_2$ and $f_{sc}$ are the forces acting on TMs and SC respectively.
	\item $f_{c2}$ and $f_{csc}$ are control forces on the second TM and the SC respectively. Since $f_{c2}$ is an internal force to the system the SC experiences a reaction force $- f_{c2}$.
	\item Dots over symbols represent time derivatives.
\end{itemize}

In the main LPF science operation mode, one TM (indicated here as $\mathrm{TM}_1$) is in free-fall and provides the reference for the other TM ($\mathrm{TM}_2$) and the SC. In order to avoid unwanted drifting, both $\mathrm{TM}_2$ and the SC are controlled to follow $\mathrm{TM}_1$. It is worth noting that the system is, by construction, symmetric and the role of the two TMs can be inverted. In order to avoid confusion, we indicate with $\mathrm{TM}_1$ the free-fall reference and with $\mathrm{TM}_2$ the actuated TM.
 
Moving to the Laplace domain, substituting for the SC dynamics and substituting for the differential coordinate $x_{\Delta}$ of $\mathrm{TM}_2$ with respect to $\mathrm{TM}_1$, the equation (\ref{eqn:lpfeqmotion1}) can be rewritten as:
\begin{equation}\label{eqn:lpfeqmotion2}
	\begin{split}
		& s^2 x_1 \omega_1^2 \left(1+\frac{m_1}{m_{sc}}\right) x_1 + \omega_2^2 \frac{m_2}{m_{sc}} x_1 + \omega_2^2 \frac{m_2}{m_{sc}} x_{\Delta} = \\
		& = \frac{f_1}{m_1} - \frac{f_{sc}}{m_{sc}} + \frac{1}{m_{sc}} H_2 o_{\Delta} - \frac{1}{m_{sc}} H_{sc} o_1 \\
		& s^2 x_{\Delta} + \left(\omega_2^2 - \omega_1^2\right) x_1 + \omega_2^2 x_{\Delta} = \\
		& = \frac{f_2}{m_2} - \frac{f_1}{m_1} + \frac{1}{m_2} H_2 o_{\Delta}.
	\end{split}
\end{equation}

Here:
\begin{itemize}
	\item $o_1$ and $o_{\Delta}$ are output displacement signals as provided by the interferometer readout system. $o_1$ is the displacement between the SC and the $\mathrm{TM}_1$. $o_{\Delta}$ is the displacement of $\mathrm{TM}_2$ relative to $\mathrm{TM}_1$.
	\item $H_2$ and $H_{sc}$ are transfer functions of the control systems on $\mathrm{TM}_2$ and SC. The force applied by the controllers is calculated on the basis of the output displacement, therefore $f_{c2} = H_2 o_{\Delta}$ and $f_{csc} = H_{sc} o_1$.
\end{itemize}

Calculations can be more easily performed if we introduce a matrix notation:
\begin{equation}\label{eqn:position}
\mathbf{x} = \left( \begin{array}{l}
 x_1  \\ 
 x_{\Delta} \\ 
 \end{array} \right)
\end{equation}
\begin{equation}\label{eqn:ifooutput}
\mathbf{o} = \left( \begin{array}{l}
 o_1  \\ 
 o_{\Delta}   \\ 
 \end{array} \right)
\end{equation}
\begin{equation}\label{eqn:forceinput}
\mathbf{f} = \left( \begin{array}{l}
 f_1  \\ 
 f_2  \\
 f_{sc}  \\ 
 \end{array} \right)
\end{equation}
\begin{equation}\label{eqn:freedynamics}
\mathbf{D} = \left( {\begin{array}{*{20}c}
   {s^2  + \omega _1^2  + \frac{m_1}{m_{sc}}\omega _1^2  + \frac{m_2}{m_{sc}}\omega _2^2 } & {\frac{m_2}{m_{sc}}\omega _2^2}  \\
   {\omega _2^2  - \omega _1^2 } & {s^2  + \omega _2^2}  \\
\end{array}} \right)
\end{equation}
\begin{equation}\label{eqn:coupling}
\mathbf{G} = \left( {\begin{array}{*{20}c}
   \frac{1}{m_1} & 0 & -\frac{1}{m_{sc}}  \\
   -\frac{1}{m_1} & \frac{1}{m_2} & 0  \\
\end{array}} \right)
\end{equation}
\begin{equation}\label{eqn:coupling}
\mathbf{C} = \left( {\begin{array}{*{20}c}
   {-\frac{1}{m_{sc}} } & { \frac{1}{m_{SC}} }  \\
   0 & {\frac{1}{m_2} }  \\
\end{array}} \right)
\end{equation}
\begin{equation}\label{eqn:controllers}
\mathbf{H} = \left( {\begin{array}{*{20}c}
   {H_{sc} } & 0  \\
   0 & {H_2 }  \\
\end{array}} \right)
\end{equation}

With the notation introduced, the equation of motion can be rewritten:
\begin{equation}\label{eqn:matrixdyn1}
	\mathbf{D} \cdot \mathbf{x} = \mathbf{G} \cdot \mathbf{f} + \mathbf{C} \cdot \mathbf{H} \cdot \mathbf{o}.
\end{equation}

Then we should consider that the output displacement $\mathbf{o}$ corresponds to the measurement of $\mathbf{x}$ provided by the sensing system:
\begin{equation}\label{eqn:radout}
	\mathbf{o} = \mathbf{S} \cdot \mathbf{x} + \mathbf{o_{rn}}.
\end{equation}

Here $\mathbf{S}$ is a $2 \times 2$ sensing matrix and $\mathbf{o_{rn}}$ is the readout noise.

Substituting equation (\ref{eqn:radout}) in equation (\ref{eqn:matrixdyn1}), the dynamics can be rewritten in terms of the known output $\mathbf{o}$:
\begin{equation}\label{eqn:matrixdyn2}
	\mathbf{D} \cdot \mathbf{S}^{-1} \cdot \mathbf{o} - \mathbf{C} \cdot \mathbf{H} \cdot \mathbf{o} = \mathbf{G} \cdot \mathbf{f} + \mathbf{D} \cdot \mathbf{S}^{-1} \cdot \mathbf{o_{rn}}.
\end{equation}

The quantity $\mathbf{G} \cdot \mathbf{f}$ represents the force-per-unit-mass acting on the test masses. Such a quantity is not directly known from the system output since the available signal is $\mathbf{o}$. From equation (\ref{eqn:matrixdyn2}) it is readily seen that an effective force-per-unit-mass acting on the TMs can be reconstructed from the knowledge of $\mathbf{o}$ if the force applied by the control system ($\mathbf{C} \cdot \mathbf{H} \cdot \mathbf{o}$) is calculated and subtracted from the reconstructed dynamics $\mathbf{D} \cdot \mathbf{S}^{-1} \cdot \mathbf{o}$.
Therefore the available quantity is $\mathbf{g} = \mathbf{G} \cdot \mathbf{f} + \mathbf{D} \cdot \mathbf{S}^{-1} \cdot \mathbf{o_{rn}}$ which contains the force-per-unit-mass acting on the TMs corrupted by the readout noise of the system.

The expected values for $\mathbf{g}$ can be obtained thanks to the dynamics reported in equation (\ref{eqn:matrixdyn2}) once the expected values for $\mathbf{f}$ and $\mathbf{o_{rn}}$ are known. The same applies to the projection of the noise spectra for $\mathbf{f}$ and $\mathbf{o_{rn}}$ to $\mathbf{g}$.

\section{Huber's norm}\label{appendix.huber}

Huber's norm \cite{Huber1973} is a way to construct a goodness of fit estimator which is more robust to outliers and non-Gausianity than the standard mean squared error (MSE). The norm is constructed as $\sum_i \rho(r_i)$, where $r_i$ are the residuals between a data series and a parametric model. The function $\rho(r_i)$ s defined as:

\begin{equation}
 \rho(r_i) =
  \begin{cases}
   \frac{1}{2} r_i^2 & \text{for } |r_i| < c \\
   c |r_i| - \frac{1}{2} c^2 & \text{for } |r_i| \geqslant c.
  \end{cases}
\end{equation}

The value of the threshold constant $c$ may depend on the given dataset, therefore some efforts must be spent to select the value of $c$ providing the most accurate results. Normalized residuals are defined as $r_i = \left(P_{WOSA}\left(f_i\right)-S_a\left(f_i;\theta\right)\right) / S_a\left(f_i;\theta\right)$ in accordance to the MSE definition in section \ref{subsect.test.noisemodelident}. With such a definition, residuals are expected to be zero mean and unitary variance in correspondence of the 'true' model $S_a\left(f_i;\theta\right)$. Different values of $c$ were tested with a Monte Carlo estimation on the first $1000$ data of the analysis reported in section 
\ref{subsect.test.noisemodelident}. Huber's norm is minimized for each MC iteration and the corresponding values of the parameter $\theta$ are stored. The histogram of $\theta$ shows a mean steadily fixed on $0.13$ for values of $c$ ranging from $0.001$ to $0.1$. Increasing the value of $c$, the distribution starts to shift toward the distribution obtained with MSE minimization. MSE and Huber distributions are practically undistinguishable for values $c > 2$. Since the 'true' value for the parameter $\theta$ is set to $0.1$, Huber's norm is performing better for values $c < 0.1$. It was then decided to use the value $c = 0.05$ for the final analysis reported in figure \ref{fig:fig5}.


\end{document}